\documentclass[final,1p,times]{elsarticle} 
\usepackage{graphicx}
\usepackage{amssymb} 
\usepackage{amsthm} 
\usepackage{lineno}

\journal{Nuclear Physics A} 

\begin{document}

\begin{frontmatter} 

% Your Title - please insert
\title{On static quark anti-quark potential at non-zero temperature}

%% Single author (and collaboration) - please insert
\author{A. Bazavov and P. Petreczky}

\address{Physics Department, Brookhaven National Laboratory, Upton, NY 11793, USA}
%% Multiple authors
%\author[auth2]{Marcus Junius Brutus}
%\address[auth1]{Somewhere, Rome}
%\address[auth2]{Somewhere else, Rome}

\begin{abstract} 
We study Wilson loops at non-zero temperature and extract
the static quark potential from them. The extracted potentials are
larger than the singlet free energies and do not show screening
for $T<190$ MeV.
\end{abstract} 

\end{frontmatter} % do not change

%% linenumbers are useful for reviewing process
%\linenumbers

\section{Introduction}
Quarkonium suppression was proposed by Matsui and Satz as a signature
of formation of deconfined medium in heavy ion collisions \cite{MS86}.
The basic idea behind this proposal is that color screening in the deconfined
medium will modify the heavy quark potential, eventually leading to the
dissolution of the heavy quarkonium states. The problem of dissolution of
quarkonium states at high temperatures could be formulated in terms of spectral
functions. Early attempts to calculate spectral functions on the lattice
have been presented in Refs. \cite{latspf}. However, extraction of the spectral functions
from lattice results on Euclidean correlation functions is quite difficult \cite{ines01}
and one should also be careful with cutoff effects in the spectral functions extracted from
the lattice \cite{freelatspf}. Furthermore, the Euclidean time correlators may not be sensitive
to the melting of the bound states at high temperatures due to the fact
that the Euclidean time extent is limited to $<1/(2T)$ (see e.g. discussions in Ref. \cite{merev}).
The effective field theory framework for heavy quark bound states, namely pQNRCD could be
a useful tool for calculating quarkonium spectral functions \cite{nora}. 
The effective field theory approach allows to rigorously define the concept of the 
static quark anti-quark potential both at zero and non-zero temperatures. One of the main
outcomes of the effective field theory analysis is the finding that at non-zero temperature
the potential has also an imaginary part, which has important consequences for the
dissolution of the quarkonium states.
While pNRQCD is 
formulated in the weak coupling framework it is possible to extend it to the non-perturbative 
regime. For example, if the binding energy is the smallest scale in the problem all
the other scales, like the thermal scales, the inverse size of the bound state and $\Lambda_{QCD}$
can be integrated out. In this case the potential should be determined non-perturbatively
and is identical to the energy of a static $Q\bar Q$ pair.
If one further neglects the dipole interactions one gets the generalization of the
simple potential model to the case of high temperatures \cite{me10}. However, one
still needs to specify the potential. In the past model
considerations based on lattice calculations of the so-called singlet free energy have been used
(see e.g. discussion in Ref. \cite{me10}). In Ref. \cite{rothkopf} it has been suggested
to extract the energy of a static $Q\bar Q$ pair using the spectral decomposition of the 
temporal Wilson loops at non-zero temperature. In this contribution we attempt to extract
the static quark anti-quark energy in 2+1 flavor QCD based on this idea.

\section{Numerical results}
In lattice QCD calculations the static $Q\bar Q$ energy is extracted from Wilson loops
$W(r,\tau)$. At large Euclidean time separations the exponential decay of the
Wilson loops is governed by the static energy or potential, $W(r,\tau)\sim \exp(-V(r)\tau)$.
More generally one can write a spectral decomposition for the Wilson loops \cite{rothkopf}
\begin{equation}
W(r,\tau)=\int_{-\infty}^{\infty} d \omega \sigma(r,\omega) e^{-\omega \tau}.
\label{spectral}
\end{equation}
At zero temperature the spectral function is proportional to $\delta(\omega-V(r))$ plus a
sum of delta functions corresponding to the excited states (hybrid potentials). At non-zero temperature the delta
function becomes a Lorentzian with the width related to the imaginary part of the potential.
In Ref. \cite{rothkopf} the potential at non-zero temperature was extracted by inverting Eq. (\ref{spectral})
via the maximum entropy method (MEM) which gives reasonably accurate determination of the real part
of the potential. At the same time it is hard to estimate the imaginary part of the potential. The problem
is similar to the problem of extracting meson spectral functions \cite{latspf,ines01}, where the width
of the bound state peaks is mostly an artifact of MEM. Another problem that appears in the calculation of the potential is
that the Wilson loops become noisy at large spatial separations $r$. To deal with this problem smeared gauge
fields are used in spatial links when constructing Wilson loops on the lattice. Alternatively,
one can fix the Coulomb gauge and calculate the correlation functions of two temporal Wilson
lines separated by distance $r$ without connecting them by spatial links \cite{milc04}. At
zero temperature where one only interested in the energy levels both choices are equally good, and
merely correspond to different choices of static meson interpolating operators.
The same should be true at non-zero temperature provided the imaginary part of the potential is
not too large.
\begin{figure}[htbp]
\begin{center}
\includegraphics[width=0.47\textwidth]{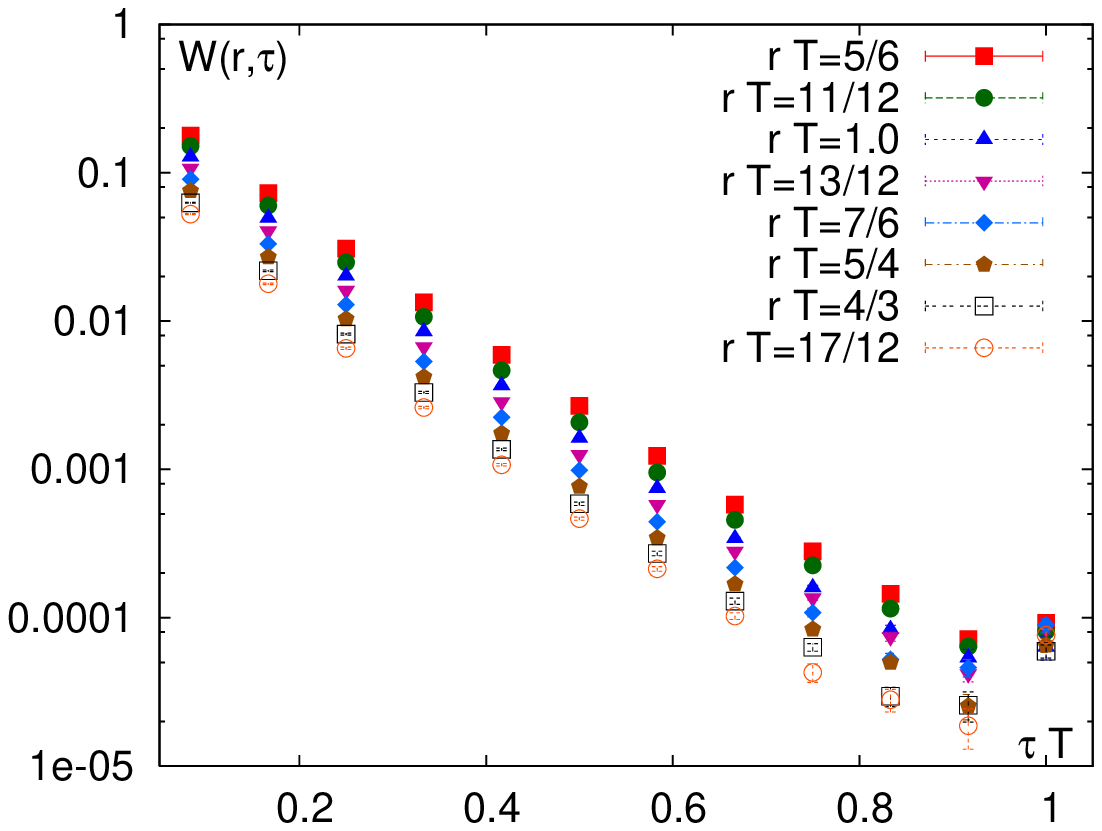}
\includegraphics[width=0.47\textwidth]{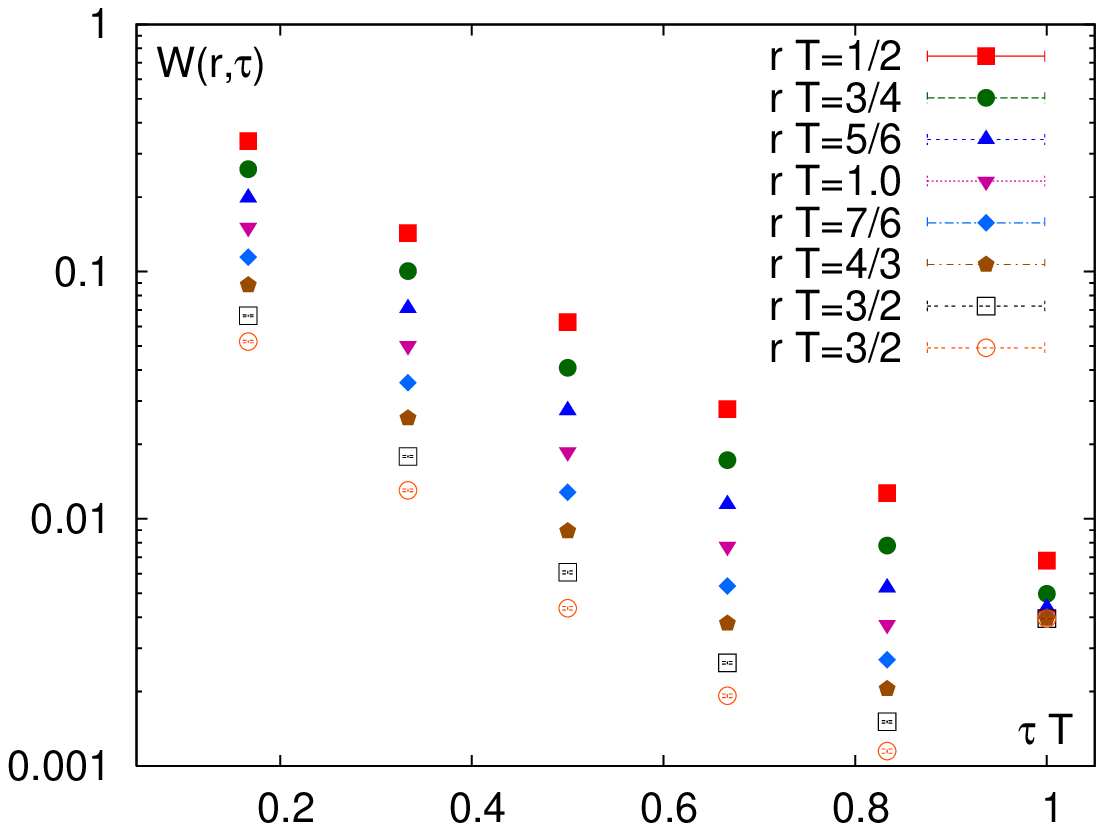}
\end{center}
\vspace*{-0.6cm}
\caption{The correlation function of Wilson lines as function of the time extent $\tau$
calculated for $48^3 \times 12$, $\beta=7.5$ (left) and $24^3 \times 6$, $\beta=6.8$ (right).}
\label{fig:w}
\vspace*{-0.3cm}
\end{figure}
We have calculated the correlation functions of temporal Wilson lines in 2+1 flavor QCD 
using the Highly Improved Staggered Quark (HISQ) action \cite{follana} and improved gauge action
with the physical value of the strange quark mass and light quark masses corresponding to pion
mass of $160$ MeV in the continuum limit.
The calculations have been performed on $48^3 \times 16$ and $48^3 \times 12$ lattices
for bare gauge coupling $\beta=10/g^2=7.5$ as well as on $24^3 \times 6$ lattices at various
gauge couplings that have been used in the study of the chiral and deconfinement transition by HotQCD \cite{tc}.
As in Ref. \cite{tc} the lattice spacing was set by the $r_1$ scale extracted from the static potential and
using the value $r_1=0.3106$ fm. We also used the renormalization constants for the Wilson line correlators
determined in Ref. \cite{tc}.
It should be noted that for $\tau T=1$ the correlator gives the so-called singlet free energy \cite{okacz02}.
The temperature dependence of the singlet free energy obtained in our 
calculations is very similar to the temperature dependence of the singlet free energy obtained  earlier 
with  the p4 action \cite{sqm09,okacz07}. 
\begin{figure}[htbp]
\begin{center}
\includegraphics[width=5.4cm]{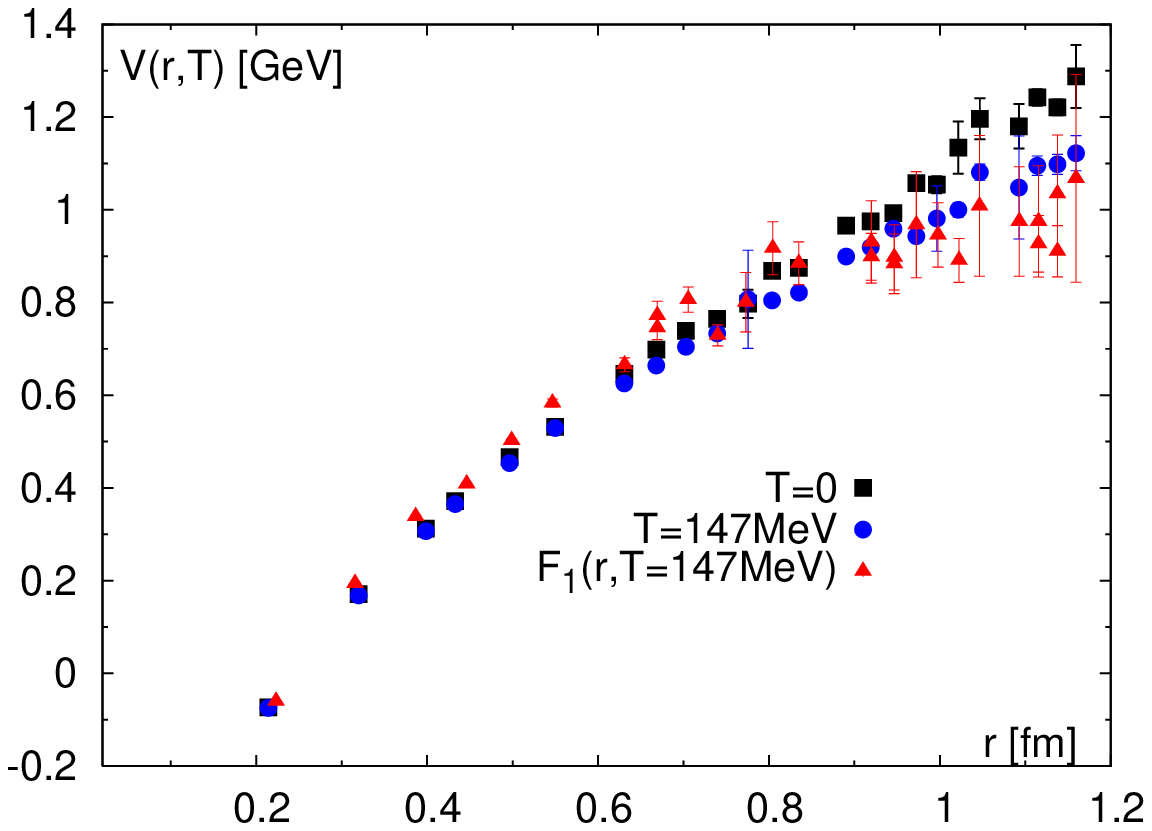}
\includegraphics[width=5.4cm]{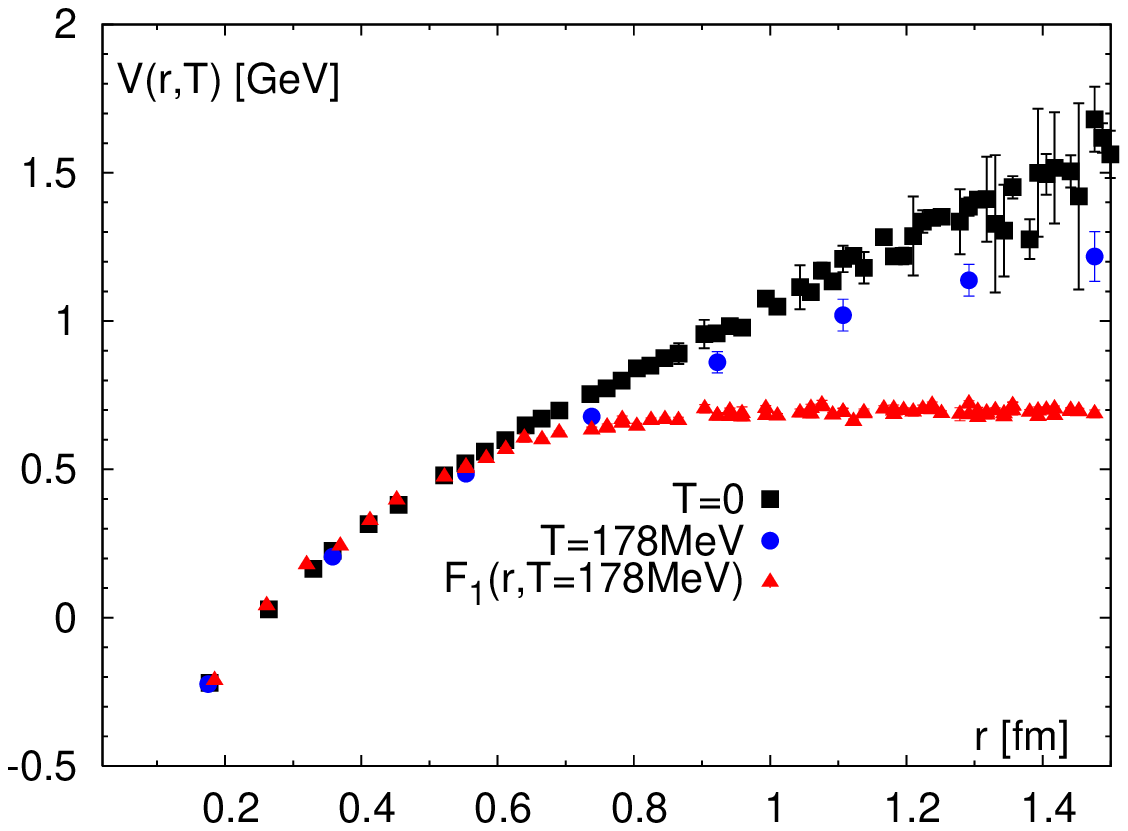}
\includegraphics[width=5.4cm]{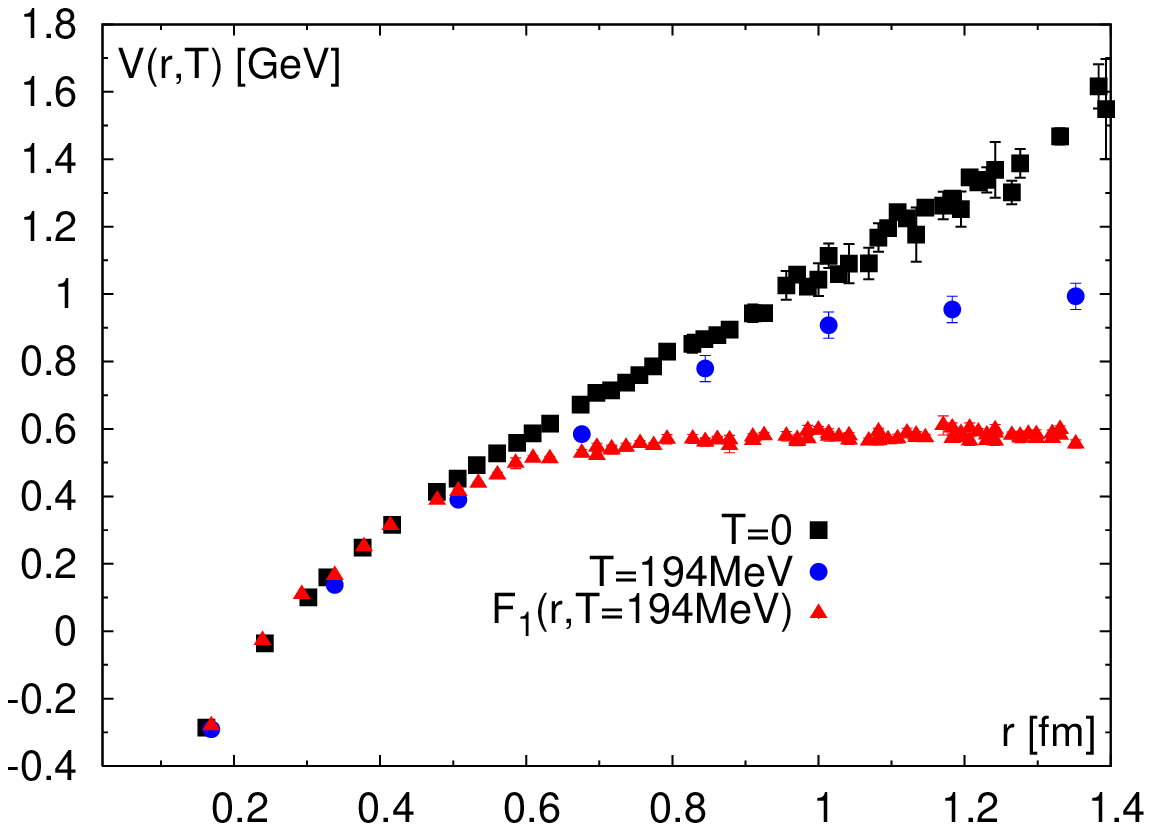}
\includegraphics[width=5.4cm]{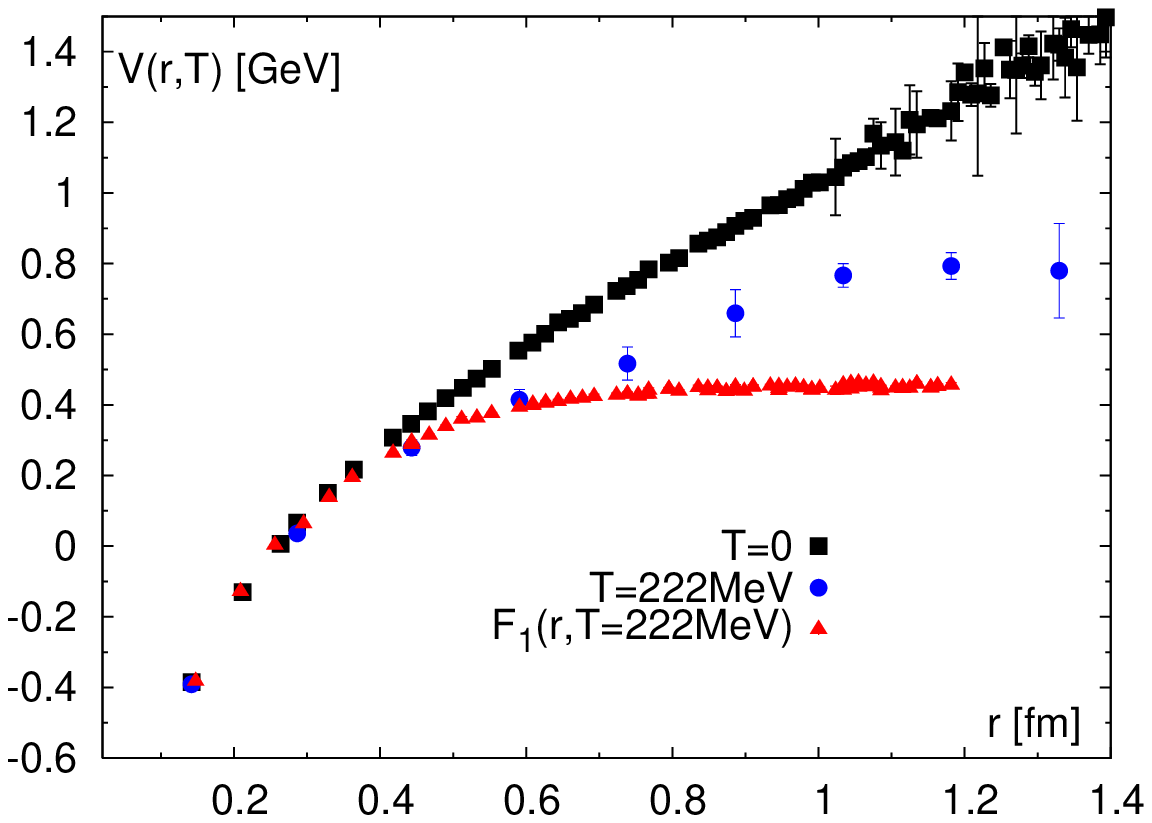}
\includegraphics[width=5.4cm]{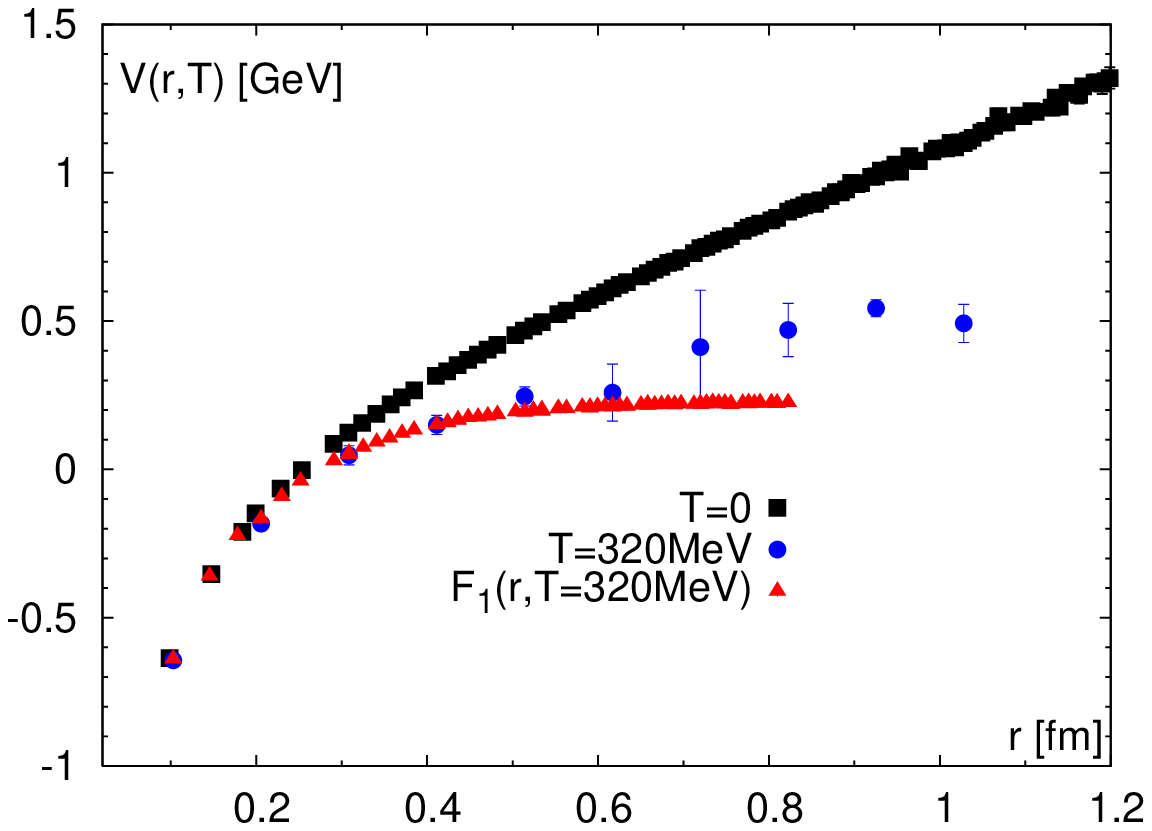}
\includegraphics[width=5.4cm]{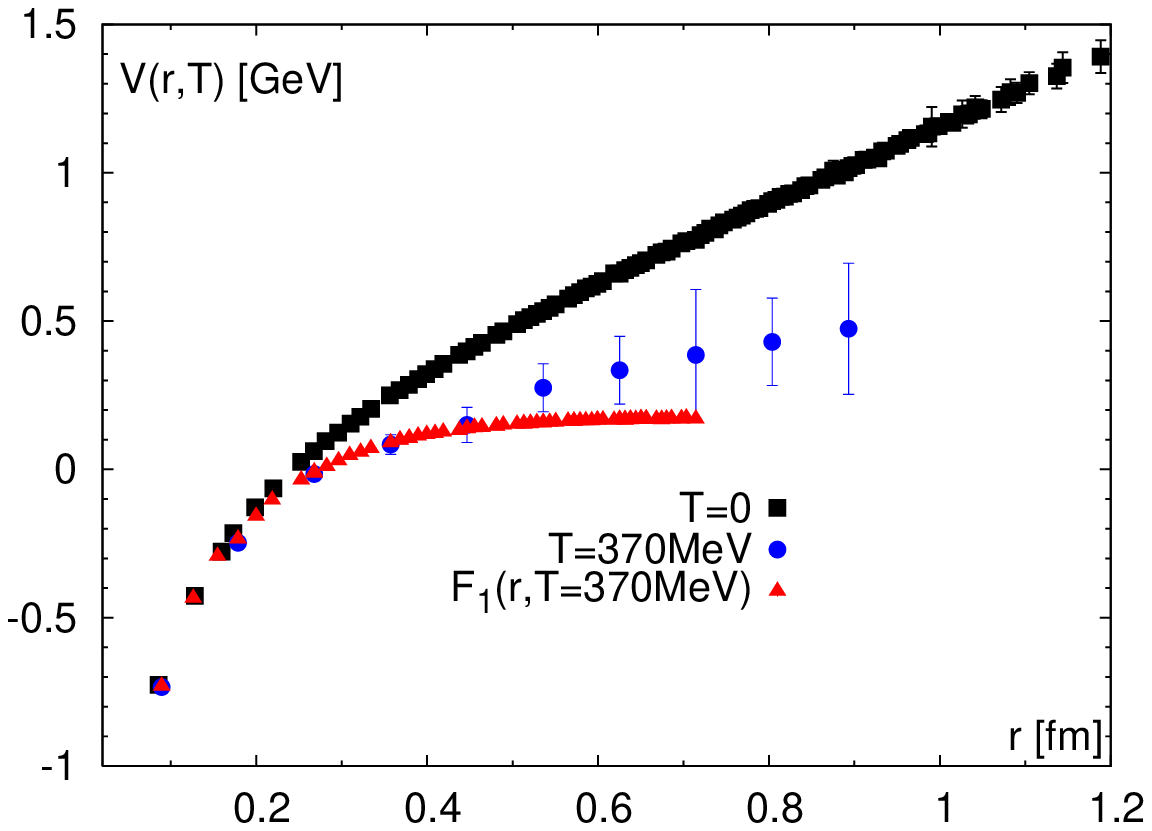}
\end{center}
\vspace*{-0.6cm}
\caption{
The static quark anti-quark potential extracted from $24^3 \times 6$ 
lattices at different temperature compared to the zero temperature result
as well as to the singlet free energy.
}
\label{fig:pot}
\vspace*{-0.2cm}
\end{figure}
In Fig. \ref{fig:w} we show our numerical results for the correlator of Wilson lines as function
of the Euclidean time extent $\tau$ for different distances calculated on $48^3 \times 12$ lattice at 
$\beta=7.5$ and $24^3 \times 6$ lattice at $\beta=6.8$. These gauge couplings correspond to temperatures
$300$ MeV and $320$ MeV, respectively. As one can see from Fig. \ref{fig:w} the $\tau$-dependence of the correlators
is consistent with simple exponential decay, except close to $\tau T \simeq 1$, where the correlators
show a slight increase. This increase is due to the contribution of a backward propagating state $\sim \exp(-E_{back}(1/T-\tau))$,
that arises from the fact that static quarks propagate in gauge field background that is periodic in $\tau$.
Similar behavior has been observed in the Wilson loop calculations at non-zero temperature 
in pure gauge theory \cite{rothkopf} as well as in full QCD calculations of bottomonium spectral
functions within the non-relativistic formulation \cite{aarts}. For the largest two values of $N_{\tau}$ considered here the backward
propagating state does not cause any problem and we get stable results for the potential by performing
single exponential fits in the $\tau$-interval around the mid-point $\tau T=1/2$. However, the results
are quite noisy for $r T>1$ in this case.
Thus to explore the potential at larger distances we use
$24^3 \times 6$ lattices for which statistical errors are small. Here the results are sensitive
to how the fits are done. We performed three type of fits. First we used only $\tau T=1/3$ and
$\tau T=1/2$ to extract the potential. Then we determined the backward propagating contribution
by performing fits with $\tau T=1$ and $\tau T=2/3$. Subtracting the  backward propagating 
contribution from the correlator we extracted the potential using $\tau T=1/2$ and $\tau T=2/3$
points. This gives our central value of the extracted potential. Finally, we performed single
exponential fits for $\tau T=1/2$ and $\tau T=2/3$ which gave us the lower value of the potential.

Our numerical results for the static quark anti-quark potential 
at different temperatures are shown in Fig. \ref{fig:pot}. The errors for
the temperature dependent potential  shown in the figure
are mostly systematics and are estimated as described above.
We also compare the static quark anti-quark potential with the  corresponding zero temperature
results as well as with the singlet free energy. 
For the lowest temperature both the potential and the singlet free energy agree with the zero temperature
result. For $T=178$MeV the singlet free energy is very different from the zero temperature potential, while
the difference is small for the finite temperature potential. Furthermore, the in-medium potential
does not show screening at this temperature. Screening effects become apparent in the potential at $T=194$MeV and 
happen at distances of about $1$fm. At the same temperature the screening effects in the singlet
free energy set in at distance of about $0.6$fm. At higher temperatures screening effects in the potential
set in at smaller and smaller distances and the difference between the potential and the singlet free energy
becomes smaller. This is expected as at very high temperatures the singlet free energy should be equal to
the potential \cite{nora}. At all temperatures the potential is larger than the singlet free energy, i.e.
it seems to approach the singlet free energy from above.  

\section{Conclusion}
We have calculated the correlation functions of Wilson lines at non-zero temperature 
for different $\tau$ and extracted the temperature dependent potential. The temperature
dependent potential is always larger than the singlet free energy, approaching it from
above. We do no see screening effects present in the potential for $T<190$MeV. The value
of the extracted temperature potential at large distance is very close to the value of
the phenomenological potential used in Ref. \cite{me10}.

\end{document}